\documentclass[9pt,conference]{IEEEtran}
\IEEEoverridecommandlockouts

\usepackage{cite}
\usepackage{amsmath,amssymb,amsfonts}
\usepackage{algorithmic}
\usepackage{graphicx}
\usepackage{textcomp}
\usepackage{xcolor}
\def\BibTeX{{\rm B\kern-.05em{\sc i\kern-.025em b}\kern-.08em
    T\kern-.1667em\lower.7ex\hbox{E}\kern-.125emX}}

\usepackage{booktabs}
\usepackage{tabularx}
\usepackage{multirow}
\let\OLDthebibliography\thebibliography
\renewcommand\thebibliography[1]{
  \OLDthebibliography{#1}
  \setlength{\parskip}{0pt}
  \setlength{\itemsep}{0pt plus 0.3ex}
}

\begin{document}

\title{Speech Recognition Rescoring with Large Speech-Text Foundation Models}
\author{\IEEEauthorblockN{Prashanth Gurunath Shivakumar, Jari Kolehmainen, Aditya Gourav, Yi Gu, Ankur Gandhe, Ariya Rastrow, Ivan Bulyko}
\IEEEauthorblockA{\textit{Amazon Science, Seattle, Washington, U.S.A} \\
\texttt{pgurunat@usc.edu, \{jkolehm,gouravag,yilegu,aggandhe,arastrow,ibbulyko\}@amazon.com}}
}

\maketitle

\begin{abstract}
Large language models (LLM) have demonstrated the ability to understand human language by leveraging large amount of text data.
    Automatic speech recognition (ASR) systems are often limited by available transcribed speech data and benefit from a second pass rescoring using LLM.
    Recently multi-modal large language models, particularly speech and text foundational models have demonstrated strong spoken language understanding.
    Speech-Text foundational models leverage large amounts of unlabelled and labelled data both in speech and text modalities to model human language.
    In this work, we propose novel techniques to use multi-modal LLM for ASR rescoring.
    We also explore discriminative training to further improve the foundational model rescoring performance.
    We demonstrate cross-modal knowledge transfer in speech-text LLM can benefit rescoring.
    Our experiments demonstrate up-to 20\% relative improvements over Whisper large ASR and up-to 15\% relative improvements over text-only LLM.
\end{abstract}


\section{Introduction}\label{sec:intro}

End-to-end speech recognition systems have made significant leaps in recognition accuracies leading to new and practical applications towards human-computer interactions.
A big chunk of the advances in end-to-end ASRs can be attributed to increase in available transcribed speech data.
This is substantiated by studies such as \cite{whisper} from OpenAI that have leveraged large amount of transcribed, labelled, speech datasets (680k hours) to advance the state-of-the-art and generalizability in multi-lingual ASR.
However, availability of transcribed speech data is always limited, due to the costs associated with acquisition and human transcription of speech.
These limitations have contributed to success of second pass rescoring with LMs.
Furthermore, with advancements in LLMs, its ability to leverage large amounts of freely available text data to demonstrate human level performance in natural language understanding benchmarks, makes second pass rescoring even more attractive.

Several recent works \cite{huang2019empirical, shin2019effective, salazar2020masked, futami2021asr, udagawa2022effect} have developed and shown effective application of second pass rescoring and associated benefits in improving recognition accuracies leveraging knowledge from LLM pre-training.
\cite{huang2019empirical} conducted empirical studies using pre-trained GPT models for re-scoring resulting in up-to 7\% relative word error rate (WER) reduction.
\cite{shin2019effective} proposed using BERT models for deriving utterance level scores for ASR re-scoring to leverage advantages from bi-directional encoding.
\cite{salazar2020masked} conducted comparisons between GPT and BERT pre-trained models and proposed mechanisms to reduce computations with BERT models.
\cite{futami2021asr} proposed to re-purpose pre-trained ELECTRA models for error detection and rescoring.
The authors also propose better pre-training and data augmentation techniques for rescoring using ELECTRA.
Authors in \cite{udagawa2022effect}, conducted a comprehensive study on application and relevance of LLMs in rescoring on state-of-the-art ASR baselines.
Their study concluded that second pass rescoring achieves consistent improvement over competitive ASR baseline models.
\cite{gandhe2020audio}, \cite{hu2020deliberation} and \cite{sainath2019two} have explored introducing audio into second pass rescoring using LSTMs and attention.

Furthermore, several works have explored novel techniques to leverage pre-training knowledge with a discriminative fine-tuning stage to optimize LLMs for ASR rescoring \cite{gurunathshivakumar23_interspeech,kolehmainen2023personalization,shivakumar2023discriminative,gu2023scaling,rescorebert}.
\cite{rescorebert} proposed a low latency framework to discriminatively finetune the BERT CLS embeddings with minimum word error rate (MWER) criteria.
\cite{shivakumar2023discriminative} proposed various methodologies for incorporating MWER criteria for fine-tuning of GPT and BERT based models.

On the other hand, generative LM technology is applied to speech using discrete audio token representations derived from audio encoders such as HuBERT \cite{lakhotia2021generative, kharitonov2022text, nguyen2023generative}.
The semantic knowledge from quantized codes provide effective way to model speech as a language using LLMs.
They have also enabled textless speech-to-speech translation \cite{lee2022textless}, speech emotion conversion \cite{kreuk2021textless} which have shown promising results, maintaining naturalistic spoken conversation and dialogs \cite{nguyen2023generative}.

More recent research have focused on modeling both speech and text tokens jointly towards cross-modal learning.
Some studies have focused on autoregressive, joint modeling on text and several speech related tasks including ASR, text-to-speech (TTS), speech-to-text translation as well as speech-to-speech translation \cite{rubenstein2023audiopalm, maiti2023voxtlm,chou2023toward}.
Others have focused on zero shot multi-modal capabilities \cite{nguyen2024spirit,zhang2023speechgpt,nachmani2023lms}.

In this paper, we propose to use multi-modal LLMs for second pass ASR rescoring.
The multi-modal LLMs are trained on a combination of unsupervised speech, text and transcribed, parallel speech-text datasets.
We propose different configurations of rescoring with speech-text foundation models and establish the advantages over text-only LLMs.
Further, we discriminatively finetune the speech-text LLMs with MWER criteria to optimize performance on rescoring.
To the best of our knowledge this is the first attempt at using multi-modal LLMs for ASR second pass rescoring.
Finally, we demonstrate cross-modal knowledge transfer can benefit rescoring in two ways: (i)
the proposed framework can leverage large amounts of un-transcribed speech data and via cross-modal knowledge transfer improve rescoring, and (ii) speech-text LLMs can associate speech information with corresponding text representations, thereby improving rescoring even when using single modality, i.e., only text tokens, during re-ranking.
Note that this work differs from \cite{gandhe2020audio,hu2020deliberation,sainath2019two} where rescoring models attend to audio representations, since they are limited to and require parallel transcribed data for training.
On the other hand, this work focuses on large scale pre-training of speech-text foundational models including unlabelled speech data and leveraging the knowledge from pre-training for rescoring.

\section{Proposed Approach}\label{sec:methodology}
\subsection{Speech-Text Foundation LLM}
A typical text-based language model, models the probability of next token given a set of tokens.
A speech-text LLM extends this paradigm to acoustic units, i.e., models the probability of the next acoustic unit given a sequence of acoustic units.
In this work, we adopt a pre-trained HuBERT encoder to derive audio representations which are quantized using k-means clustering into discrete audio tokens.
Text tokens are derived using a sentence piece models.
A decoder-only transformer architecture is used for causal language modeling on both text and audio tokens.
Given a sequence of tokens $Z = {z_1, z_2, \ldots z_T}$, the next token prediction tasks model:
\begin{equation}
P_{LM}(Z) = \prod_{i=1}^T P(z_i|z_{i-1},\ldots z_1)
\end{equation}
where
$z_i \in {V_{txt} \cup V_{speech}}$ is the multi-modal token sequence, $V_{txt}$ is the vocabulary corresponding to text tokens, $y_i\in V_{txt}$, $V_{speech}$ is the vocabulary corresponding to speech tokens, $x_i \in V_{speech}$ (see Table~\ref{tab:data_format}) and $P_{LM}(z_{T+1})$ is the likelihood.

Similar to \cite{maiti2023voxtlm}, we use unsupervised training on audio-only and text-only tokens as speech continuation and text continuation tasks respectively.
Additionally, we also utilize parallel, transcribed speech data for joint modeling of text and speech by creating two versions of concatenated multi-modal token sequence, i.e., text followed by speech and vice-versa.
The data format used for pre-training is listed in Table~\ref{tab:data_format}.

\begin{table}
\begin{tabular}{l}
\midrule
$\langle$text-bos$\rangle$ $y_1, \ldots y_T$ $\langle$eos$\rangle$ \\
$\langle$speech-bos$\rangle$ $x_1, \ldots x_L$ $\langle$eos$\rangle$ \\
$\langle$text-bos$\rangle$ $y_1, \ldots y_T$ $\langle$speech-bos$\rangle$ $x_1, \ldots x_L$ $\langle$eos$\rangle$ \\
$\langle$speech-bos$\rangle$ $x_1, \ldots x_L$ $\langle$text-bos$\rangle$ $y_1, \ldots y_T$ $\langle$eos$\rangle$ \\
\midrule
\end{tabular}
\caption{Speech-Text LLM pre-training data format}\label{tab:data_format}
\end{table}

\subsection{ASR Rescoring}
\subsubsection{Likelihood-based Rescoring}
Typical text LM rescoring comprises computing likelihood scores from the model for each of the n-best hypothesis from ASR and interpolating with the first pass scores to re-rank the hypothesis:
\begin{equation}
s_i = \mathrm{log} P_{LM}(y_i) + \lambda log P_{AM}(a|y_i)
\end{equation}
where $P_{AM}(a|y_i)$ is the sequence probability of the 1st pass given an audio sequence, $a$, for $i^{th}$ ASR hypothesis, $y_i$, $P_{LM}(y_i)$ is the likelihood from the 2nd pass rescoring LM and $\lambda$ is the interpolation weight.

\subsubsection{Speech-Text LLM Rescoring}
Multi-modal LM allows us to condition the likelihood of the n-best hypothesis from ASR on the sequence of audio tokens.
\begin{equation}\label{eq:mmlm_interp}
s_i = log P_{MMLM}(z_i) + \lambda log P_{AM}(a|y_i)
\end{equation}
where $P_{MMLM}(z_i)$ is the likelihood from the multi-modal LM, $z_i$ is the multi-modal sequence containing both audio and text tokens corresponding to the $i^{th}$ ASR hypothesis.

In this work, we explore two ways to construct the multi-modal sequence for rescoring: (i) speech-first, i.e., speech precedes the text token sequence (see row 4 in Table~\ref{tab:data_format}): 
\begin{equation}
\begin{split}
P_{MMLM}(x_i) = \prod_{t=1}^{T_i} P(y_{i,t}|y_{i,t-1}\ldots y_{i,1}, x_{L}, \ldots x_1) \\[-10pt]
\prod_{j=1}^{L} P(x_{j}|x_{j-1}\ldots x_1)
\end{split}
\end{equation}
and (ii) text-first, i.e., text precedes the audio token sequence (see row 3 in Table~\ref{tab:data_format}):
\begin{equation}
\begin{split}
P_{MMLM}(x_i) =  \prod_{j=1}^{L} P(x_{j}|x_{j-1}\ldots x_1, y_{i,T},\ldots y_{i,1}) \\[-10pt] \prod_{t=1}^{T_i} P(y_{i,t}|y_{i,t-1}\ldots y_{i,1}) \\
\end{split}
\end{equation}
where $T_i$ is the length of ASR hypothesis $y_i$, and $L$ is the length of audio token sequence $X$, corresponding to the input audio.

\subsection{Discriminative Rescoring}
LLMs are sub-optimal for re-ranking since they are fundamentally trained to optimize for the next token prediction task.
Several works \cite{gurunathshivakumar23_interspeech,kolehmainen2023personalization,shivakumar2023discriminative,gu2023scaling,rescorebert} have shown benefits in optimizing the LLM to minimize the expected word edit distance for second pass rescoring.
We propose to use discriminative fine-tuning for Speech-Text LLM to further optimize towards better rescoring.
The MWER criterion for multi-modal LLM can be expressed as:
\begin{equation}
L_{mwer}(a, y^{*}) = \sum_{i=1}^N P(x_i|a) \epsilon (y_i, y^{*})
\end{equation}
where $N$ is the top-N hypothesis from ASR first-pass, $y^{*}$ is the ground-truth transcription, $\epsilon$ is the edit distance function and $P(x_i|a)$ is the n-best posterior probability given by:
\begin{equation}
P(x_i|a) = \frac{e^{s_i}}{\sum_{j=1}^N e^{s_j}}
\end{equation}
where $s_i$ is derived from Eq.~\eqref{eq:mmlm_interp} in case of multi-modal LLM.

\begin{table*}[!t]
\centering
\begin{tabularx}{1\textwidth}{X|lll|lll}
\toprule
Model & \multicolumn{3}{c|}{Whisper large v2 \cite{whisper}} & \multicolumn{3}{c}{Conformer RNN-T} \\
Dataset & LS test-clean & LS test-other & Tedlium & LS test-clean & LS test-other & Tedlium\\
\midrule
First pass & 2.26\% & 5.32\% & 4.87\% & Baseline & Baseline & Baseline \\
\midrule
330M text LM \cite{salazar2020masked} & 2.26\% (0\%) & 5.3\% (0.75\%) & 4.64\% (4.72\%) & (4.12\%) & (2.46\%) & (2.81\%) \\
\hspace{10mm} + MWER \cite{shivakumar2023discriminative} & 1.94\% (14.12\%) & 4.89\% (8.43\%) & 4.48\% (8.01\%) & (9.28\%) & (8.79\%) & (8.2\%) \\
330M speech-text LM (TF) & 2.08\% (7.96\%) & 4.95\% (7.3\%) & 4.43\% (9.03\%) & (5.15\%) & (3.51\%) & (7.34\%) \\
330M speech-text LM (SF) & 1.92\% (15.04\%) & 4.68\% (12.36\%) & 4.48\% (8.01\%) & (11.34\%) & (7.73\%) & (4.97\%) \\
\hspace{10mm} + MWER & 1.86\% (17.7\%) & 4.65\% (12.92\%) & 4.49\% (7.8\%) & (13.4\%) & (9.67\%) & (7.78\%) \\
7B text LM \cite{salazar2020masked} & 2.25\% (0.44\%) & 5.27\% (1.31\%) & 4.58\% (5.95\%) & (6.87\%) & (1.41\%) & (2.59\%) \\
\hspace{10mm} + MWER \cite{shivakumar2023discriminative} & 2.15\% (4.87\%) & 5.12\% (4.12\%) & 4.66\% (4.31\%) & (10.31\%) & (11.78\%) & (7.78\%) \\
7B speech-text LM (SF) & 1.94\% (14.16\%) & 4.62\% (13.48\%) & 4.34\% (10.88\%)& \textbf{(19.93\%)} & (13.36\%) & (12.1\%) \\
\hspace{10mm} + MWER & \textbf{1.85\% (18.14\%)} & \textbf{4.55\% (14.79\%)} & \textbf{3.16\% (35.11\%)} & (15.94\%) & \textbf{(15.99\%)} & \textbf{(14.47\%)} \\
\midrule
Yu et. al. (2024) \cite{yu2024connecting} & 2.1\% (7.08\%) & 5.0\% (6.02\%) & - & - & - & - \\
SALMONN \cite{tang2023salmonn} & 2.1\% (7.08\%) & 4.9\% (7.89\%) & - & - & - & - \\
Oracle WER & 1.57\% (30.53\%) & 3.86\% (27.72\%) & 2.53\% (48.05\%) & (42.96\%) & (37.73\%) & (38.88\%) \\
\bottomrule
\end{tabularx}
\vspace{1mm}
\caption{Experimental Results: Word error rate. Numbers within parenthesis is relative improvements with respect to the first pass. SF: speech-first followed by text; TF: text-first followed by speech.}\label{tab:results}
\end{table*}

\section{Data and Experimental Setup}\label{sec:setup}

\subsection{Experimental Setup}
\begin{table}[b]
\begin{tabular}{lll}
\toprule
Dataset & Hours & Modality \\
\midrule
Librispeech \cite{panayotov2015librispeech} & 960 & Speech, Text \\
Librilight \cite{kahn2020libri} & 60k & Speech \\
Multi-lingual librispeech \cite{pratap2020mls} & 50k & Speech, Text \\
People Speech \cite{galvez2021people} & 30k & Speech, Text \\
CoVOST2 \cite{wang2020covost} & 3500 & Speech, Text \\
TEDLIUM-3 \cite{hernandez2018ted} & 452 & Speech, Text \\
De-identified Internal data & 791 & Speech, Text \\
\bottomrule
\end{tabular}
\vspace{1mm}
\caption{Speech-Text Foundation Model Data Setup}\label{tab:data}
\end{table}

Our setup uses a HuBERT audio encoder that is trained on multi-lingual datasets as in ~\cite{lee2022textless} to derive the audio tokens.
The HuBERT operates at the rate of 50Hz.
A k-means clustering model is trained on the same data as the HuBERT using 2000 centroids.
The HuBERT audio encoder is frozen throughout the training of the speech-text LLM for all our experiments.


We employ two different sized LLMs for our experiments differing in the model size (i) 330M parameters, and (ii) 7B parameters.
The smaller, 330M model, is similar to OPT architecture \cite{zhang2022opt} with 24 hidden layers with a size of 1024, 16 attention heads, intermediate dimension of 4096, embedding dimension of 512 and vocabulary of 50466 (inclusive of 2000 audio tokens).
The bigger, 7B model, is similar to Llama architecture \cite{touvron2023llama} with 32 hidden layers with a size of 4096, 32 attention heads, embedding dimension of 4096 and vocabulary of 52001 (inclusive of 2000 audio tokens).
Both the models are pre-trained on text-only data, and the vocabulary is extended to include the speech-tokens for subsequent training with mixed modalities, following findings from \cite{hassid2024textually}.
Both the models use sentence-piece tokenizers with a vocabulary of approximately 50k text-tokens and 2000 audio tokens.
The models are trained with Adam optimizer using exponential decay learning rate scheduler with learning rate of 1e-5 and 1000 warmup steps.
Our 330M speech-text model achieves a sWUGGY score of 63.7\% and sBLIMP of 55.5\%.
The 7B speech-text LLM achieves a sWUGGY score of 67.7\% and sBLIMP of 55.5\%.

\subsection{ASR rescoring}
We employ two first pass systems: (i) open-sourced whisper (large v2) from OpenAI \cite{whisper}, (ii) conformer based RNN-T model, and present the results on multiple open-source datasets to demonstrate the generalization capability of the proposed technique.
The conformer RNN-T model is trained on a combination of internal and public speech datasets.
The WER and Oracle WER of the first pass models are listed in Table~\ref{tab:results}.
For the internal model, we present relative improvements with respect to the first pass.
Note the absolute WER of the internal first pass model is better than Whisper.
For all our experiments, we use Top-10 hypothesis for rescoring purposes similar to \cite{gurunathshivakumar23_interspeech,shivakumar2023discriminative}.
The re-scoring setup is identical between the two first pass systems.
In case of MWER fine-tuning, the optimal checkpoints are picked on validation sets and the results are presented on unseen held-out test-sets.

\subsection{Data}
The speech-text LLM was initially pre-trained with large text corpus.
The 330M model is based on the pre-training setup as described in \cite{zhang2022opt} followed by multi-modal training on multi-lingual Libri-Speech \cite{pratap2020mls}.
The 7B model was pre-trained using RedPajama \cite{together2023redpajama}.
Table~\ref{tab:data} lists the datasets that were employed during multi-modal training for the 7B model.
Approximately 145k hours of publicly available speech corpora is used and 800 hours of de-identified internal data.
This accounts for approximately 26.1B speech tokens.
Additionally, for evaluations on out-of-domain datasets, we use Wall Street Journal (WSJ) \cite{paul1992design}, Common-Voice (English) \cite{ardila2020common} and AMI meeting corpus \cite{carletta2005ami}.

For all the MWER training experiments, 50k hours of multi-lingual Librispeech is adopted for both training and validation.
The interpolation weights ($\lambda$ in Eq.~\eqref{eq:mmlm_interp}) for all the experiments is estimated on the validation partition of multi-lingual Librispeech comprising of approximately 1.2k utterances.


\section{Results}\label{sec:results}
Table~\ref{tab:results} lists the experimental results.
We provide the absolute WER numbers in case of Whisper (large v2) and relative improvements for Conformer RNN-T.
Note, in the case of Whisper, any small discrepancies in the WER with respect to \cite{whisper} can be attributed to differences in transcript normalization.
Firstly, we observe that typical log-likelihood based text LM rescoring with 330M model improves over the first pass by less than 5\% relative.
The proposed multi-modal rescoring with 330M speech-text LLM provides substantial improvements over text-only LLM on both the first pass Whisper (up-to 15\% relative) and Conformer RNN-T models (up-to 6\% relative).
This suggests that the proposed method can exploit, through autoregressive language modeling of audio tokens, important information relevant for ASR rescoring.

In our experiments, we find that constructing sequences with audio-tokens first, followed by text is superior (see results comparing SF versus TF in Table~\ref{tab:results}).
However, we note that regardless of the order, the audio-tokens are helpful in rescoring in comparison to text-only LLMs which further confirms the effectiveness of the proposed technique.

Furthermore, we observe that the discriminative training with MWER criteria always improves the WER in all cases, both on text-only LLM and speech-text LLM especially on Librispeech.
Note that the Tedlium is out-of-domain as far as the MWER fine-tuning is concerned.
Interestingly, the speech-text LLM without MWER training can achieve lower WER in comparison to MWER-trained text LLMs, in most cases.
Adding discriminative training to speech-text LLM helps further extend the advantage over text LLMs.

Experiments with 7B model paints similar picture to that of the 330M model, thereby demonstrating the effectiveness of the proposed technique over text-only LLMs extending to different scales.
We observe that log-likelihood based rescoring with text-only LLMs scales poorly with the model size.
However, we find speech-text LLMs to scale better especially in the case of Conformer RNN-T.
Next, with discriminative tuning, the 7B model provides significant improvements suggesting that discriminative fine-tuning scales well with the model size.
Overall, we obtain the best results with the discriminatively trained 7B speech-text model.

We also provide comparisons with other approaches that combine Whisper large and 7B LLMs for ASR (see bottom 3 rows in Table~\ref{tab:results}). We find that our proposed rescoring framework provides significant advantages and a viable alternative.

Finally, the results generally depicts similar trends with both first pass models and different datasets.
This shows the generalizability of the proposed technique over different first pass systems and over variety of data sets.

\begin{table}[t]
\centering
\begin{tabular}{lllll}
\toprule
\multirow{2}{*}{Model} & \multirow{2}{*}{WSJ} & CV & AMI & AMI \\
& & (en) & (IHM) & (SDM) \\
\midrule
7B text LLM & 0.56\% & -4.88\% & 4.82\% & 5.19\% \\
\hfill + MWER & 7.04\% & 0.45\% & -15.74\% & -7.45\%\\
7B speech-text LLM & 8.52\% & -7.7\% & 11.12\% & 9.7\% \\
\hfill + MWER & 7.04\% & 3.98\% & 11.12\% & 9.77\% \\
\bottomrule
\end{tabular}
\vspace{1mm}
\caption{Results with Conformer RNN-T on out-of-domain datasets. Relative WER-reduction over the first-pass. Negative numbers indicate degradations over first-pass.}\label{tab:ood}
\end{table}

\begin{table}[b]
\centering
\begin{tabular}{llll}
\toprule
\multirow{2}{*}{Model} & \multicolumn{2}{c}{Librispeech} & \multirow{2}{*}{Tedlium} \\
& test-clean &  test-other & \\
\midrule
7B text LLM & Baseline & Baseline & Baseline \\
\hfill + MWER & 6.45\% & 10.68\% & 5.11\% \\
\midrule
7B speech-text LLM & 0.12\% & 0.89\% & 0.67\% \\
\hfill + MWER & 9.68\% & 11.39\% & 7.33\% \\
\bottomrule
\end{tabular}
\vspace{1mm}
\caption{Cross-modal experimental results on Conformer RNN-T comparing text-only rescoring with text-only LLM versus Speech-text LLM. Relative improvements with respect to text LLM rescoring.}\label{tab:cross_modal}
\end{table}

\subsection{Out-of-Domain Experiments}
Next, we design experiments to assess the impact of proposed rescoring models on out-of-domain datasets.
Table~\ref{tab:ood} lists the results using Conformer RNN-T as first pass on WSJ, common-voice (English), AMI-IHM and AMI-SDM datasets.
Note that none of these datasets are used either in training or as validation/tuning.
We observe that the text-only LLMs do not help on WSJ and common-voice datasets, while giving approximately 5\% improvements on AMI corpuses.
However, the speech-text LLM provides substantial improvements (almost double than that of text-only LLMs) on all datasets except common-voice.
We note that the MWER finetuning with text-only LLMs can have detrimental effect on out-of-domain data that have largely different acoustic characteristics (AMI), especially when they are not included during MWER fine-tuning.
However, multi-modal LLMs can counteract such effects with the ability to attend to audio.
Again on out-of-domain datasets, we observe best results with the proposed speech-text LLMs with up-to 7\% WER reduction and find that multi-modal LLMs generalize and perform better compared to their text-only counterparts.

\subsection{Cross Modal Experiments}
One of the advantages of multi-modal LLM, speech-text foundational models, is that the knowledge from one modality can be transferred to the other.
For example, the model can associate a likelihood of a sequence in one modality by inherently modeling it with corresponding sequence in another modality, when modeled jointly.
To investigate this cross-modal knowledge transfer, we perform two set of experiments:\\
\textbf{Experiment 1:} Text-only rescoring on (i) text-LLM, and compare it with text-only rescoring on (ii) speech-text LLM. 
The experimental results are listed in Table~\ref{tab:cross_modal} with 7B models and Conformer RNN-T first pass.
We observe that even when using only text modality for rescoring, the speech-text LLM gives lower WER (up-to 0.8\% relative word-error reduction).
This cross-modal knowledge infusion is better exploited when further fine-tuning the LLM with MWER loss with text modality giving improvements of up-to 3.4\% relative to its text-LLM counterpart.
This suggests that during pre-training, there is some knowledge transfer from audio sequence into the text-sequence.\\
\textbf{Experiment 2:} We target a scenario of domain adaptation where transcripts are not available and hypothesize that by modeling speech-only data with speech-text LLM, can benefit multi-modal re-scoring.
We consider two 330M models trained (i) without Tedlium, and (ii) with Tedlium speech-only data.
The rescoring results is presented in Table~\ref{tab:cross_modal_tedlium}.
We observe improvements double on target Tedlium dataset with the speech-text model after speech-only adaptation on Tedlium.
This confirms that the proposed technique can leverage knowledge between the two modalities efficiently and overall enhance rescoring performance.

\begin{table}[t]
\centering
\begin{tabular}{llll}
\toprule
\multirow{2}{*}{Model} & \multicolumn{2}{c}{Librispeech} & \multirow{2}{*}{Tedlium} \\
& test-clean &  test-other & \\
\midrule
330M speech-text & \multirow{2}{*}{7.53\%} & \multirow{2}{*}{5.41\%} & \multirow{2}{*}{2.22\%} \\
(no-Tedlium) & & \\
\midrule
330M speech-text & \multirow{2}{*}{8.96\%} & \multirow{2}{*}{4.14\%} & \multirow{2}{*}{4.22\%} \\
+ Tedlium audio-only & & \\
\bottomrule
\end{tabular}
\vspace{1mm}
\caption{Cross-modal experimental results on Conformer RNN-T comparing with and without target domain (Tedlium) audio-only data during speech-text foundation model training. Relative improvements with respect to text LLM (row 2 in Table~\ref{tab:results})}\label{tab:cross_modal_tedlium}
\end{table}

\section{Conclusion}\label{sec:conclusion}
In this work, we propose a second pass rescoring system for speech recognition based on multi-modal LLM.
The speech-text LLM is trained on unlabelled text and speech data in addition to parallel transcribed speech data.
We demonstrate the benefits of rescoring the ASR hypothesis with combination of speech and text sequences.
We also explore different ordering of the two modalities and their effects on rescoring performance.
Additionally, discriminative training using MWER for speech-text model is applied to further improve rescoring.
Experiments are setup to demonstrate effectiveness on wide-variety of in-domain and out-of-domain datasets.
We also show cross-modal knowledge transfer with speech-text LLM in application to rescoring.

\bibliographystyle{IEEEtran}
\scriptsize{\bibliography{mybib}}

\end{document}